\documentclass[journal]{IEEEtran}

\usepackage{fancyhdr}
\usepackage{latexsym}

\usepackage{epsfig}
\usepackage{hyperref}

\usepackage{pstricks}

\usepackage{shadow,epsf,amsthm,amssymb,amsmath}

\usepackage{enumerate}
\usepackage{graphicx}
\usepackage{setspace}                           

\usepackage{lipsum}

\newtheorem{definitionenv}{Definition}
\newtheorem{lemmaenv}[definitionenv]{Lemma}
\newtheorem{theoremenv}[definitionenv]{Theorem}
\newtheorem{corollaryenv}[definitionenv]{Corollary}
\newtheorem{propositionenv}[definitionenv]{Proposition}
\newtheorem{conjectureenv}[definitionenv]{Conjecture}
\newtheorem{exampleenv}{Example}
\newtheorem{app-lemmaenv}[section]{Lemma}

\newenvironment{definition}{\begin{definitionenv}\rm}{\end{definitionenv}}
\newenvironment{lemma}{\begin{lemmaenv}\rm}{\end{lemmaenv}}
\newenvironment{theorem}{\begin{theoremenv}\rm}{\end{theoremenv}}
\newenvironment{corollary}{\begin{corollaryenv}\rm}{\end{corollaryenv}}
\newenvironment{example}{\begin{exampleenv}\rm}{\end{exampleenv}}
\newenvironment{proposition}{\begin{propositionenv}\rm}{\end{propositionenv}}
\newenvironment{conjecture}{\begin{conjectureenv}\rm}{\end{conjectureenv}}
\newenvironment{app-lemma}{\begin{app-lemmaenv}\rm}{\end{app-lemmaenv}}

\newcommand{\bd}{\begin{definition}}
\newcommand{\ed}{\end{definition}}
\newcommand{\bl}{\begin{lemma}}
\newcommand{\el}{\end{lemma}}
\newcommand{\elp}{\hspace*{\fill} $\Box$
                 \end{lemma}}
\newcommand{\bt}{\begin{theorem}}
\newcommand{\et}{\end{theorem}}
\newcommand{\etp}{\hspace*{\fill} $\Box$
                 \end{theorem}}
\newcommand{\bc}{\begin{corollary}}
\newcommand{\ec}{\end{corollary}}
\newcommand{\ecp}{\hspace*{\fill} $\Box$
                 \end{corollary}}
\newcommand{\bcj}{\begin{conjecture}}
\newcommand{\ecj}{\end{conjecture}}

\newcommand{\be}{\begin{example}}
\newcommand{\ee}{\end{example}}
\newcommand{\eep}{\hspace*{\fill} $\Box$
                 \end{example}}
\newcommand{\bp}{\begin{proposition}}
\newcommand{\ep}{\end{proposition}}
\newcommand{\epp}{
                 \end{proposition}}

\newcommand{\wt}[1]{\mbox{wt}({#1})}

\newcommand{\ket}[1]{|#1\rangle}

\begin{document}

\title{Duality in Entanglement-Assisted\\
 Quantum Error Correction}
\author{Ching-Yi Lai, Todd A. Brun, and Mark M. Wilde\thanks{
TAB and CYL were supported in part by NSF Grant CCF-0830801.
MMW\ acknowledges the support of the MDEIE (Qu\'{e}bec) PSR-SIIRI
international collaboration grant.

        C.-Y. Lai and T. A. Brun are with the
        Communication Sciences Institute, Electrical Engineering Department,
University of Southern California, Los Angeles, California, USA  90089.
(e-mail:laiching@usc.edu and tbrun@usc.edu)

        M. M. Wilde is a postdoctoral fellow with the School of Computer
Science, McGill University, Montreal, Quebec, Canada H3A 2A7. (e-mail: mwilde@gmail.com)


This work was presented at the 14th Workshop on Quantum Information Processing in January 2011 (QIP2011) in Singapore.

       }
       }

\date{\today}

\maketitle

\begin{abstract}
The dual of an entanglement-assisted quantum error-correcting (EAQEC) code
is defined from the orthogonal group of a simplified stabilizer group.
From the Poisson summation formula, this duality leads to the MacWilliams identities and linear programming bounds for EAQEC codes.
We establish a table of upper and lower bounds on
the minimum distance of any maximal-entanglement EAQEC code with length up to 15 channel qubits.

\end{abstract}


\section{Introduction}

The theory of quantum error correction underpins the practical realization of quantum computation and quantum communication \cite{Shor95,Ste96a,Ste96b,EM96,BDSW96,KL97}.
Quantum stabilizer codes are an extensively analyzed class of quantum error-correcting codes
because their encoding, decoding, and recovery are straightforward to describe using their algebraic properties \cite{CRSS97,CRSS98,Got97,NC00}.

Entanglement-assisted quantum error correction  is a paradigm in which the sender and receiver share
entanglement before quantum communication begins \cite{BDM06}. An
$[[n,k,d;c]]$ EAQEC code encodes $k$ information qubits into $n$
channel qubits with the help of $c$ pairs of maximally-entangled
Bell states. The code can correct up to $\lfloor \frac{d-1}{2}
\rfloor$ errors acting on the $n$ channel qubits, where  $d$ is
the minimum distance of the code. Standard stabilizer codes are a
special case of EAQEC codes with $c=0$, and we use the notation
$[[n, k, d]]$ for such codes.

Bowen proposed the first EAQEC code \cite{Bowen02}, which is equivalent to the well-known five-qubit code \cite{LMPZ96}.
Fattal {\it et al}.~established a technique for handling entanglement in the stabilizer formalism \cite{FCYBC04}.
Brun, Devetak, and Hsieh then devised the entanglement-assisted stabilizer formalism and showed how to transform any $[n,k,d]$ classical quaternary code\footnote{An $[n, k, d]$ classical linear code over a certain field encodes $k$
information digits into $n$ digits, where  $d$ is its minimum distance.} into an $[[n,2k-n+c,d;c]]$ EAQEC code, where $c$ depends on the properties of the classical code~\cite{WB-2008-77}.
Lai and Brun further explored the properties of EAQEC codes and proposed an optimization method to find optimal EAQEC codes that cannot be obtained
by the aforementioned construction \cite{LB10}.
By optimal, we mean that $d$ is the highest achievable minimum distance for given parameters $n$, $k$, and $c$.

In classical coding theory, a well-established notion is that of a
dual code. Suppose that $\mathcal{C}$ is an $[n,k]$ linear code
over an arbitrary field GF$(q)$ with a  generator
matrix $G$ and  a corresponding parity check matrix $H$ such that $ HG^{T}=0. $
The dual code of $\mathcal{C}$ is the $[n,k'=n-k]$ linear code
$\mathcal{C}^{\perp}$ with  $H$ as a generator matrix and $G$ as a
parity check matrix.
The dimensions of the code $\mathcal{C}$ and
its dual code $\mathcal{C}^{\perp}$  satisfy the relation
$    k+k'=n.$
It is well known that the MacWilliams identity  gives a relationship
between the weight enumerator of $\mathcal{C}$ and the weight
enumerator of its dual code $\mathcal{C}^{\perp }$ \cite{MS77},
which can be used to determine the minimum distance of the dual
code  $\mathcal{C}^{\perp }$, given the weight enumerator of
$\mathcal{C}$.

The MacWilliams identity for quantum codes connects the weight
enumerator of a classical quaternary self-orthogonal code associated with the quantum code to the weight enumerator of its dual code \cite{SL96,Rains96b,Rains98,AL99}.
This leads to the linear programming bounds (upper bound) on the minimum distance of quantum codes.
We will show that this type of the MacWilliams identity for quantum stabilizer codes can be directly obtained by applying the Poisson summation formula from the theory of orthogonal groups.
However, the orthogonal group of a stabilizer group with respect to the symplectic inner product (which will be defined later) does not define another quantum stabilizer code.  So this is not a duality between codes in the usual quantum case.

In this paper, we define a notion of duality in entanglement-assisted quantum error correction based on the theory of orthogonal groups,
and this notion of duality bears more similarity to the classical notion of duality because the orthogonal group of
an entanglement-assisted code forms a nontrivial entanglement-assisted quantum code. We then show how the quantum analog of
the MacWilliams identity and the linear programming bound for EAQEC codes follow in a natural way.
We apply the EAQEC code constructions in \cite{BDM06,LB10,LB11a} to find good EAQEC codes with maximal entanglement for $n\leq 15$.
Combining the results of the linear programming bounds, we give a table of upper and lower bounds on the highest achievable
minimum distance of any maximal-entanglement EAQEC code\footnote{One might wonder why we are considering
EAQEC codes that exploit the maximum amount of entanglement possible, given that noiseless entanglement could be expensive in practice.
But there is good reason for doing so. The so-called ``father" protocol is a random entanglement-assisted quantum code
\cite{arx2005dev,PhysRevLett.93.230504},
and it achieves the entanglement-assisted quantum capacity
of a depolarizing channel (the entanglement-assisted hashing bound \cite{PhysRevLett.83.3081,Bowen02}) by exploiting maximal entanglement. Furthermore,
there is numerical evidence that maximal-entanglement turbo codes come within a few dB of achieving
the entanglement-assisted hashing bound \cite{WH10}.} for $n\leq 15$.
%
%


%
%
We organize this paper as follows.
We first review some basics of entanglement-assisted quantum codes in the following section.
In Section \ref{sec:dual_code}, we define the dual of an EAQEC code.
The MacWilliams identity for EAQEC codes and the linear programming bound for EAQEC codes are derived in Section \ref{sec:MacWilliams}, followed by
a table of upper and lower bounds on the minimum distance of any EAQEC code with maximal entanglement and $n\leq 15$.
The final section concludes with a summary and future questions.

\section{Review of EAQEC Codes}

We begin with some notation.
The Pauli matrices
$$I=\begin{bmatrix}1 &0\\0&1\end{bmatrix}, X=\begin{bmatrix}0 &1\\1&0\end{bmatrix}, Y=\begin{bmatrix}0 &-i\\i&0\end{bmatrix}, Z=\begin{bmatrix}1 &0\\0&-1\end{bmatrix}$$
form a basis of the space of linear operators on a two dimensional single-qubit state space $\mathcal{H}$.
Let
$$\mathcal{G}_n=\{e M_1\otimes \cdots \otimes M_n:M_j\in \{I, X, Y, Z\}, e\in \{\pm 1, \pm i \} \} $$
be the $n$-fold Pauli group.
We use the subscript $X_j$, $Y_j$, or $Z_j$ to denote a Pauli operator on qubit number $j$.
We define $X^{u}=\prod_{i:u_i=1} X_{i} $ for some binary $n$-tuple $u=(u_1\cdots u_n)$ and similarly $Z^{v}= \prod_{j:v_j=1} Z_{j} $ for some binary $n$-tuple $v=(v_1\cdots v_n)$.
Any element $g=e M_1\otimes \cdots \otimes M_n\in \mathcal{G}_n$ can be expressed as $g= e' X^{u}Z^{v}$
for some $e'\in \{\pm 1, \pm i \}$ and two two binary $n$-tuples $u$ and $v$.
The weight $\wt{g}$ of $g$ is the number of  $M_j$'s that are not equal to the identity operator $I$.
Since the overall phase of a quantum state is not important,
we consider the quotient of the Pauli group by its center
$\bar{\mathcal{G}}_n= \mathcal{G}_n/\{\pm 1, \pm i\}$,
which is an Abelian group and can be generated by a set of $2n$ independent generators.
For $g_1=  X^{u_1}Z^{v_1}$, $g_2= X^{u_2}Z^{v_2} \in \bar{\mathcal{G}}_n$,  the symplectic inner product $*$ in $ \bar{\mathcal{G}}_n $ is defined by
\[
g_1*g_2= u_1\cdot v_2+u_2\cdot v_1 \mod{ 2},
\]
where $\cdot$ is the usual  inner product for binary $n$-tuples.
Note that $*$ is commutative.
We define a map $\phi: \mathcal{G}_n\rightarrow \bar{\mathcal{G}}_n$ by
$\phi\left(e X^{u}Z^{v}\right)=  X^{u}Z^{v}$.
For $g,h\in \mathcal{G}_n$, $\phi(g)*\phi(h)=0$ if they commute,
 and  $\phi(g)*\phi(h)=1$, otherwise.
The orthogonal group of a subgroup $V$ of $\bar{\mathcal{G}}_n$ with respect to $*$ is
$$V^{\perp}= \{ g\in \bar{\mathcal{G}}_n: g*h=0, \forall h\in V \}.$$
For example, consider a stabilizer group $\mathcal{S}$, which is an Abelian subgroup of $\mathcal{G}_n$ and does
not contain the negative identity operator $-I$.
Then the orthogonal group of $\phi(\mathcal{S})$ is $(\phi(\mathcal{S}))^{\perp}= \phi( \mathcal{N}(\mathcal{S}) )$,
where $\mathcal{N}(\mathcal{S})$ is the normalizer group of $\mathcal{S}$.

An $[[n,k,d]]$ stabilizer code is a $2^k$-dimensional subspace of the $n$-qubit Hilbert space $\mathcal{H}^{\otimes n}$,
which is the joint $+1$-eigenspace of $n-k$ independent generators of a stabilizer subgroup $\mathcal{S}$ of $\bar{\mathcal{G}}_n$.
The minimum distance $d$ is the minimum weight of any element in $\phi(\mathcal{N}(\mathcal{S}))\setminus \phi(\mathcal{S})$.

In the scheme of EAQEC codes \cite{BDM06,LB10},
Alice and Bob share $c$  maximally-entangled pairs
$|\Phi_{+}\rangle_{AB}=\frac{1}{\sqrt{2}}\left(  |00\rangle+|11\rangle\right)$.
Suppose Alice tries to send a $k$-qubit state $|\phi\rangle$ to Bob through a noisy channel,
using an additional $n-k-c$ ancilla qubits in the state $\ket{0}$.
We assume that Bob's qubits suffer no errors since they do not pass through the noisy channel.
Let $J=(J^i, J^e, J^a)$ be the set of positions of the information qubits, the entangled pairs, and the ancilla qubits on Alice's side, respectively.
For example, if the initial state is $\ket{\phi}\ket{\Phi_+}_{AB}^{\otimes c}\ket{0}^{\otimes n-k-c}$,
we have $J^i=\{1,\cdots, k\}$,  $J^e=\{k+1,\cdots, k+c\}$, and $J^a=\{k+c+1,\cdots, n\}$.
Then Alice applies a \emph{Clifford encoder} $U$ on her $n$ qubits to protect the information qubits.
A Clifford encoder is a unitary operator that maps elements $\bar{\mathcal{G}}_n$ to elements of $\bar{\mathcal{G}}_n$ under unitary conjugation.
An  $\left[  \left[  n,k,d;c\right]  \right]$  EAQEC code is defined by the pair $(U, J)$, where $d$ is the minimum distance and will be defined later.
For convenience, let $g_j= UZ_j U^{\dag}$ and $h_j=UX_jU^{\dag}$ for $j=1, \cdots, n$ in $\bar{\mathcal{G}}_n$.
The encoded state associated with $(U,J)$ 
has a set of stabilizer generators
\begin{align*}
\{&g_{J_1^e}^A\otimes Z_{1}^B,\cdots,g_{J_{c}^e}^A\otimes Z_{c}^B, \\
  &h_{J_1^e}^A\otimes X_{1}^B,\cdots,h_{J_c^e}^A\otimes X_{c}^B,  \\
  &g_{J_1^a}^A\otimes I^B,\cdots, g_{J_{n-k-c}^a}^A\otimes I^B\}
\end{align*}
 in $\bar{\mathcal{G}}_{n+c},$
where the superscript $A$ or $B$ indicates that the operator acts on the qubits of Alice or Bob, respectively,
and $J_j^{x}$ denotes the $j$-th element in the set $J^{x}$.
Since Bob's qubits are error-free, we only consider the operators on Alice's qubits.
The simplified stabilizer subgroup $\mathcal{S}'$  associated with the pair $(U, J)$ of $\bar{\mathcal{G}}_n$ is 
\[\mathcal{S}'= \langle g_{J_1^e},\cdots,g_{J_{c}^e},  h_{J_1^e},\cdots,h_{J_c^e},    g_{J_1^a},\cdots, g_{J_{n-k-c}^a}\rangle.\]
Note that the commutation relations are as follows:
\begin{align}
&g_i*g_j=0\mbox{ for all $i$ and $j$}, \label{eq:commutation_1}\\
&h_i*h_j=0\mbox{ for all $i$ and $j$}, \label{eq:commutation_2}\\
&g_i*h_j=0 \mbox{ for $i\neq j$}, \label{eq:commutation_3}\\
&g_i*h_i=1 \mbox{ for all $i$}. \label{eq:commutation_4}
\end{align}
We say that $g_j$ and $h_j$ are  \emph{symplectic partners} for $j\in J^i\cup J^e$.
The logical subgroup $\mathcal{L}$ associated with the pair $(U, J)$ of the encoded state is
$$\mathcal{L}= \langle g_{j}, h_j ,j\in J^i  \rangle.$$
The symplectic subgroup $\mathcal{S}_S$ associated with the pair $(U, J)$ of $\mathcal{S}'$  is the subgroup generated by the $c$ pairs of symplectic partners of $\mathcal{S}'$:
$$\mathcal{S}_S= \langle g_{j}, h_j ,j\in J^e\rangle.$$
The isotropic subgroup $\mathcal{S}_I$ associated with the pair $(U, J)$ of $\mathcal{S}'$ is the subgroup generated by the generators $g_i$ of $\mathcal{S}'$.
Therefore $g_i*g=0$ for all $g$ in $\mathcal{S}'$:
$$\mathcal{S}_I =\langle g_{j},j\in J^a \rangle.$$
Notice that  $\mathcal{S}'= \mathcal{S}_S\times \mathcal{S}_I$  in  $\bar{\mathcal{G}}_n$.
The minimum distance $d$ of the EAQEC code is the minimum weight of any element in
$\mathcal{S}'^{\perp}\setminus \mathcal{S}_I$.

\section{Duality in EAQEC Codes}\label{sec:dual_code}
Observe that the orthogonal group of $\mathcal{S}'=\mathcal{S}_S\times \mathcal{S}_I$  associated with the pair $(U, J=(J^i,J^e,J^a))$  in $\bar{\mathcal{G}}_n$ is
$\mathcal{L}\times \mathcal{S}_I$. That is,
\[
\mathcal{L}\times \mathcal{S}_I=(\mathcal{S}_S\times \mathcal{S}_I)^{\perp}.
\]
We can define another EAQEC code with logical subgroup $\mathcal{S}_S$, symplectic subgroup $\mathcal{L}$ and isotropic subgroup $\mathcal{S}_I$  associated  to the pair $(U,J'=(J^e,J^i,J^a))$.

The number of a set of independent generators of $ \mathcal{S}_S\times \mathcal{S}_I $ is  $K= 2c+(n-k-c)=n-k+c$,
and the number of a set of independent generators of its orthogonal group  $\mathcal{L}\times \mathcal{S}_I$ is  $K'=2k+ (n-k-c)=n+k-c $.
These parameters satisfy the following relation:
\[ K+K'=2n=N,\]
where $N$ is the number of a set of independent generators of the full Pauli group $\bar{\mathcal{G}}_n$.
This equation is parallel to
the classical duality between a code and its dual code,
which motivates the definition of the dual code of an EAQEC code as follows.
\bd
The dual of an $[[n,k,d;c]]$ EAQEC code, defined by a simplified stabilizer group $\mathcal{S}'=\mathcal{S}_S\times \mathcal{S}_I$ and a logical group $\mathcal{L}$ associated with the pair $(U, J=(J^i,J^e,J^a))$,
is the $[[n,c,d';k]]$ EAQEC code associated with the pair $(U, J'=(J^e,J^i,J^a))$ where  $\mathcal{L}\times \mathcal{S}_I$ is the simplified stabilizer group and $\mathcal{S}_S$  is the logical group for some minimum distance $d'$.
\ed

When $c=n-k$, we call such a code a {\it maximal-entanglement} EAQEC code.
In this case, $\mathcal{S}_I$ is the trivial group that contains only the identity, and
the simplified stabilizer group is $\mathcal{S}_S$.
Its dual code is a maximal-entanglement EAQEC code  defined by the logical group $\mathcal{L}$.

When $c=0$, the code is a standard stabilizer code,
with a stabilizer group $\mathcal{S}= \mathcal{S}_I=\langle g_{j}, j\in J^a\rangle$, and a logical group
$\mathcal{L}=\langle g_j, h_j, j\in J^i \rangle$.
$\mathcal{S}_S$ is the trivial group in this case.
The simplified stabilizer group  $\mathcal{L}$ defines an $[[n,
0,d',k]]$ EAQEC code---that is, a single entangled stabilizer state
that encodes no information.


\section{The MacWilliams Identity and the Linear Programming Bounds} \label{sec:MacWilliams}
The MacWilliams identity for general quantum codes can be obtained from the general theory of classical additive codes as indicated in \cite{CRSS98} or by applying the Poisson summation formula  from the theory of orthogonal groups \cite{Knapp06}.
\bt \label{thm:MacWilliamsIdentityOrthogonalGroup}
Suppose $W_V(x,y)= \sum_{w=0}^n B_w x^{n-w} y^w$ and $W_{V^{\perp}}(x,y)= \sum_{w'=0}^n A_{w'} x^{n-w'} y^{w'}$
are the weight enumerators of  a subgroup $V$ of $\bar{\mathcal{G}}_n$ and  its orthogonal group $V^{\perp}$  in $\bar{\mathcal{G}}_n$.
Then
\begin{align}
W_{V}(x,y) =\frac{1}{|V^{\perp}|} W_{V^{\perp}}(x+3y,x-y). \label{eq:MacWilliamsIdentityOrthogonalGroup}
\end{align}
or equivalently
\begin{align}
B_w=\frac{1}{|V^{\perp}|} \sum_{w'=0}^n \label{eq:MacWilliamsIdentityCoeff}
P_w(w',n) A_{w'},
\ \mbox{ for $w=0, \cdots, n$},
\end{align}
where $
P_w(w',n)= \sum_{u=0}^{w}(-1)^u 3^{w-u} {w'\choose u}{n-w'\choose
w-u}
$
is the Krawtchouk polynomial \cite{MS77}.
\et

Applying Theorem \ref{thm:MacWilliamsIdentityOrthogonalGroup} to  the simplified stabilizer group $\mathcal{S}_S\times \mathcal{S}_I$ and the isotropic subgroup $\mathcal{S}_I$, respectively,
we obtain the MacWilliams Identity for EAQEC codes.
\bc \label{thm:GeneralQuantumMacWilliamsEAQEC}
The MacWilliams identities for EAQEC codes are as follows:
\begin{equation}
W_{\mathcal{L}\times \mathcal{S}_I}(x,y)=\frac{1}{|\mathcal{S}_S\times \mathcal{S}_I|}  W_{\mathcal{S}_S\times \mathcal{S}_I}(x+3y,x-y), \label{eq:MIEAQEC1}
\end{equation}
\begin{equation}
W_{ \mathcal{S}_I}(x,y)=\frac{1}{|\mathcal{L}\times \mathcal{S}_S\times \mathcal{S}_I|}  W_{\mathcal{L}\times \mathcal{S}_S\times \mathcal{S}_I}(x+3y,x-y).\label{eq:MI EAQEC2}
\end{equation}

\ec

The significance of the MacWilliams identities  is that linear
programming techniques  can be applied to find upper bounds on the
minimum distance of EAQEC codes.
For an $[[n,k,d; n-k]]$ EAQEC code, $\mathcal{S}_I$ is trivial and the minimum distance is the minimum weight of any element
in the logical subgroup $\mathcal{L}$.
We must have $B_w=0$ for $w=1,\cdots, d-1$.
If we cannot find any solutions to an integer program with the following constraints:
\begin{align*}
&A_0=B_0=1;\\
&A_w\geq 0, B_w \geq 0, \mbox{ for $w=1, \cdots, n$};\\
&A_w\leq |\mathcal{S}_S|, B_w \leq |\mathcal{L}|, \mbox{ for $w=1, \cdots, n$};\\
&\sum_{w=0}^n A_w= |\mathcal{S}_S|, \sum_{w=0}^n B_w =|\mathcal{L}|;\\
&B_w= \frac{1}{|\mathcal{S}_S|} \sum_{w'=0}^n P_w(w',n) A_{w'}, \mbox{ for $w=0,\cdots, n$};\\
&B_w=0,   \mbox{ for $w=1, \cdots, d-1$};
\end{align*}
for a certain $d$, this result implies that there is no $[[n,k,d; n-k]]$ EAQEC code. If
$d^*$ is the smallest of such $d$'s, then $d^*-1$ is an upper bound
on the minimum distance of an $[[n,k,d;n-k]]$ EAQEC code. This bound is
called the linear programming bound for EAQEC codes.

For $0<c<n-k$, both $\mathcal{S}_I$ and $\mathcal{S}_S$ are nontrivial.
The minimum distance is the minimum weight of any element in $\mathcal{S}_I\times \mathcal{L} \setminus \mathcal{S}_I$.
We need constraints on both the weight enumerators of $\mathcal{S}_I\times \mathcal{S}_S$ and $\mathcal{S}_I$
from equations (\ref{eq:MIEAQEC1})  and (\ref{eq:MI EAQEC2}).

For $c=0$,
 $V^{\perp}$ is the stabilizer group $\mathcal{S}$, $V$ is the normalizer group of $\mathcal{S}$ and (\ref{eq:MacWilliamsIdentityOrthogonalGroup}) gives the MacWilliams Identity for stabilizer codes \cite{SL96,Got97}.

%


Now we can establish a table of upper and lower bounds on the minimum distance of maximal-entanglement EAQEC codes for $n\leq 15$.
The upper bounds for $n\leq 15$ and $k\geq 2$ are from the linear programming bound,
which is generally tighter than
the singleton bound \cite{BDM06}
and the Hamming bound for nondegenerate EAQEC codes \cite{Bowen02}.
The linear programming bounds are not necessarily tight, however.  In some cases, they can be improved by other arguments.  For instance, it can be proved
that $[[n,1,n;n-1]]$ and $[[n,n-1,2;1]]$ EAQEC codes do not exist for even $n$ \cite{LBW10}.

Lai and Brun proposed a construction of $[[n, 1,n;n-1]]$
entanglement-assisted (EA) repetition codes for $n$ odd in \cite{LB10}.
By slightly modifying that construction, we construct $[[n, 1,n-1;n-1]]$ EA repetition
codes for $n$ even \cite{LBW10}, which  are optimal.

The following codes are obtained by applying the  the EAQEC code construction from classical codes in \cite{BDM06}:
$[[7,2,5;5]]$, $[[9,4,5;5]]$, $[[9,5,4;4]]$, $[[10,4,6;6]]$, $[[11,5,6;6]]$, $[[11,4,6;7]]$, $[[11,6,5;5]]$,
$[[12,2,9;10]]$, $[[12,8,4;4]]$, $[[12,5,6;7]]$,
$[[13,2,10;11]]$, $[[13,3,9;10]]$,  $[[13,6,6;7]]$, $[[14,7,6;7]]$, $[[14,8,5;6]]$,
$[[15,9,5;6]]$, $[[15,8,6;7]]$.
The following codes are from  the circulant code construction  in \cite{LB10}:
 $[[7, 3, 4; 4]]$, $[[8, 2, 6; 6]]$, $[[10, 3, 6; 7]]$, $[[11,3,7;8]]$, $[[15,5,8;10]]$, $[[15,6,7;9]]$.
The following codes are obtained by transforming standard stabilizer codes into EAQEC codes in \cite{LB11a}:
$[[6,2,4;5]]$,  $[[8,4,4;4]]$, $[[9,6,3;3]]$, $[[10,6,4;4]]$, $[[10,7,3;3]]$,
$[[11,8,3;3]]$, $[[12,6,5;6]]$, $[[12,7,4;5]]$, $[[12,9,3;3]]$,
$[[13,5,6;8]]$, $[[13,9,4;4]]$, $[[13,10,3;3]]$,
 $[[14,11,3;3]]$,
$[[15,4,8;11]]$, $[[15,10,4;5]]$.
These codes give lower bounds on the achievable distance for many values of $n$ and $k$.

If an $[[n,k,d;c]]$ code exists, it can be shown that both an $[[n+1,k,d;c+1]]$ and  an $[[n,k-1,d'\geq d;c+1]]$ code exist \cite{LBW10},
which proves the existence of the following codes:
$[[14,3,9;11]]$, $[[14,9,4;5]]$, 
$[[13,8,4;5]]$,  $[[14,10,3;4]]$, $[[14,6,6;8]]$,  codes.
We used MAGMA \cite{MAGMA} to find the optimal quantum stabilizer codes, and then
applied the encoding optimization algorithm in \cite{LB10} to obtain the other lower bounds.

\begin{table}
  \begin{tabular}{|c|c|c|c|c|c|c|c|}
\hline
      $n\backslash k$&    $1$& $2$ & $3$&4&5&6&7\\
      \hline
      3 &  3& 2  &  &&&&\\
      4 &  3& 2-3&1 &&&& \\
      5 &  5& 3-4&   2-3&2&&&\\
      6 &  5& 4&  3-4 &2&1&&\\
      7 & 7 &5&4&3&2&2&\\
      8 & 7& 6&5&4 &3 & 2&1\\
      9 & 9& 6-7& 5-6& 5& 4& 3& 2 \\
      10 & 9&7-8&6-7&6&4-5&4&3\\
      11 & 11&8&7-8&6-7&6&5&3-4\\
      12 & 11&9&7-8&6-7&6-7&5-6&4-5\\
      13 & 13 &10&9&6-8&6-7&6-7&4-6\\
      14 & 13&10-11&9-10&7-9&6-8&6-7&6-7\\
      15 & 15 &11-12&9-11&8-10&8-9&7-8&6-7\\
\hline
%
%
%
\hline
      $n\backslash k$& 8&9&  10& $11$& $12$&13&14 \\
      \hline
      9  &2&& &&&&\\
      10 &2&1 &&&&&\\
      11 &3&2& 2&& &&\\
      12 &4&3&2&1&&& \\
      13 &4-5&4&3&2&2&& \\
      14 &5-6&4-5&3-4&3&2&1& \\
      15 &6-7&5-6&4&3-4&2-3&2&2 \\
\hline
 \end{tabular}

   \caption{Upper and lower bounds on the minimum distance of any $[[n,k,d;n-k]]$
   maximal-entanglement EAQEC codes.
   } \label{tb:Bounds}

\end{table}

\section{Discussion} \label{sec:Discussion}

In this paper, we defined the dual code of an EAQEC code and derived the MacWilliams
identities for EAQEC codes. Based on these identities, we found a linear
programming bound on the minimum distance of a EAQEC code.
We provided a table of upper and lower bounds on
the minimum distance of maximal-entanglement EAQEC codes for $n\leq 15$.
Most lower bounds in Table \ref{tb:Bounds}
are from the optimization algorithm \cite{LB10}.
To make the bounds in Table \ref{tb:Bounds} tighter, 
we need to consider other code constructions to raise the lower bounds.
We also plan to explore the existence of
other $[[n,k,d;n-k]]$ codes to decrease the upper bound.
 Similar tables for EAQEC codes with $0<c<n-k$ can be constructed by the same
techniques.

Rains introduced the idea of  the \emph{shadow} enumerator of a quantum stabilizer code \cite{Rains96}, which
can be related to the weight enumerator of the stabilizer group similar to the MacWilliams identity.
This relation provides additional constraints on the linear programming problem and
can improve the linear programming bound for quantum codes.
To introduce the ``shadow enumerator" of an EAQEC code may be a potential way to improve on the linear programming bound.

We are indebted to an anonymous referee and Associate Editor Jean-Pierre Tillich for constructive comments on our manuscript.
MMW acknowledges useful discussions with Omar Fawzi and Jan Florjanczyk.

\end{document}